\documentclass[preprint,10pt]{revtex4}
\usepackage{graphicx}

\begin{document}

\title[Network Entropy applied to perturbed cell states]{Network Entropy measures applied to different systemic perturbations of cell basal state}
\author{G. Menichetti\,$^{1}$, G. Bianconi\,$^{2}$,  E. Giampieri\,$^1$, G. Castellani\,$^1$ and D. Remondini\,$^1$}
\affiliation{$^{1}$Department of Physics and Astronomy, Bologna University, Viale B. Pichat 6/2 40127 Bologna, Italy\\
$^{2}$School of Mathematical Sciences, Queen Mary University of London, London E1 4NS, United Kingdom}

\maketitle

\section*{Abstract}
We characterize different cell states, related to cancer and ageing phenotypes, by a measure of entropy of network ensembles, integrating gene expression values and protein interaction networks. The entropy measure estimates the parameter space available to the network ensemble, that can be interpreted as the level of plasticity of the system for high entropy values (the ability to change its internal parameters, e.g. in response to environmental stimuli), or as a fine tuning of the parameters (that restricts the range of possible parameter values) in the opposite case.
This approach can be applied at different scales, from whole cell to single biological functions, by defining appropriate subnetworks based on a priori biological knowledge.

Network entropy was able to characterize specific pathological (relapsing/metastatic samples) or extreme longevity (successful ageing samples, including centenarians) conditions.
For the cancer ``case-control" situation (primary tumour samples which developed metastasis/relapsed or not) we observed a global reduction of entropy for the most aggressive tumor samples, both at a whole-genome level and for a statistically significant selection of genes, as if a specific genetic profiling was necessary to obtain such specific conditions.
The ageing dataset was characterized globally by a smaller entropy for the extremely aged group, and by a non-monotonic trend for the genes significantly changing over time, coherent with the previous whole-genome studies.
Moreover, scaling the analysis from whole-genome to specific biological pathways allowed a deeper understanding of the cell processes involved.\\
This measure was able to characterize different experimental designs (case-control and time series studies) with a high accuracy and robustness. 
In our analysis we used specific network features (degree sequence, subnetwork structure and distance between gene profiles) to obtain informations at different biological scales, providing a novel point of view for the integration of experimental transcriptomic data and a priori biological knowledge, but the entropy measure can also highlight other aspects of the biological systems studied depending on the constraints introduced in the model (e.g. community structures). 

Matlab and Python code for the calculation of Network Entropy $S_{NE}$ (given an adjacency matrix and a distance matrix) are available upon request.

\section{Introduction}

Biological systems can be seen as complex units that translate genomic information into phenotypes, encoded in the proteome/in\-te\-rac\-to\-me \cite{Pagel,DeLasRivas10}.
A useful approach applied in several contexts, from biological to social and artificial systems, is to describe such systems as networks, with the system elements as nodes, and the relationships between them as edges \cite{barabasi,AlmArkin03,Strogatz01}.
An outstanding class of biological networks comprises the protein-protein interaction networks (PPI \cite{vidal,pathwaycommons,BossiLehner09}): edges in these networks describe interactions between proteins that are part of the same physical complex or post-translational modifications mediating signal transduction flows. Networks of interacting proteins can be thought  as the phenotypes of sets of functionally linked genes. 
Topological studies have shown that PPI networks have peculiar features, characterized by most proteins having small degree and with a small yet significant number of highly connected nodes, called hubs. Since these hubs have been shown to encode for essential genes \cite{barabasi}, highlighting the topological properties of these networks may carry important biological information.\\
In our study we are interested in the integration between the transcriptomic and the interactomic data, thus the statistical properties of integrated PPI-mRNA expression networks seem to be good observables to investigate systemic pathologies such as cancer and ageing \cite{Severini,ageingInteractome}. This approach can be more informative than analyzing gene expression data on its own. Indeed, integrative PPI-mRNA expression studies have helped to tease out relevant patterns of expression variation in the contextual framework of signaling pathways and protein complexes \cite{Pagel,westSciRep12,vanWieringen11}.\\
Following the recent developments in statistical mechanics of complex networks we have the chance to build up a thorough biological network model. Thanks to some suitable constraints encoding the most relevant network features, we can evaluate the information content of biological structures, and moreover, we can apply specific methods for time-dependent and time-independent data \cite{Bianconi,BianconiPNAS}.\\
Our integrated PPI-mRNA network approach relies on the theory of network ensembles with given topology (degree sequence) and metrics (distance between the nodes). The PPI structure is embedded in the network topology, while mRNA expression data define the distance values between network elements. This amount of information is translated into constraints on the number of interactions per each protein and on the number of interactions per each distance bin (where the distance between two proteins depends on the associated gene expression values).\\
We studied two biological phenomena that encode different landscapes of cellular perturbation, namely cancer and ageing in humans, and whose datasets were characterized by a different experimental design (a case-control study and a time series analysis).\\
Our approach offers a new perspective to the study of such phenomena, highlighting a more systemic behavior of the cell beyond single-element analysis. In particular, we study the effect of these perturbations at several scales, from a whole-cell point of view to the single biological pathways characterizing the main cell processes (like metabolism and signaling).

\section{Methods}

\subsection*{Datasets}

\subsubsection*{PPI network}
We used the protein-protein interaction network extracted from the Pathway Commons database (www.pathwaycommons.org) regarding \textit{Homo Sapiens} proteins. The initial PPI network contained 11604 nodes and 420601 links: after self-interaction and redundant annotation removal we obtained a giant component of 11394 nodes and 420516 links, that we considered for our analysis. For each gene expression dataset, we considered a subset of this PPI network obtained by the intersection of the protein IDs with the gene annotations of the microarray probes. Moreover, several induced subnetworks (described in the Subsection below and in the Supplementary Material) were extracted by intersecting these networks with the pathway annotations obtained by the KEGG database (Kyoto Encyclopaedia of Genes and Genomes, www.genome.jp/kegg).

\subsubsection*{Cancer dataset}
The analysis has been performed onto three datasets considered in a previous work that used an entropy-like measure \cite{Severini}, by downloading the normalized data from GEO Omnibus (www.ncbi.nlm.nih.gov/geo) or Arrayexpress (www.ebi.ac.uk/arrayexpress). All data sets were profiled with the Affymetrix U133 Plus 2 platform, and referred to primary invasive breast cancer samples that were followed for the evidence of relapse or metastasis. We refer to them as ``Wang'' (GSE2034 \cite{Wang}) , ``Loi'' (GSE2990 \cite{Loi2,Loi}), ``Frid'' (E-TABM-158 \cite{Frid}). ``Wang" dataset consisted of 286 samples (primary tumor biopsies taken at the early stage), 107 classified as ``relapsing" and 179 as ``not relapsing" after follow-up. ``Loi" and ``Frid" datasets were collections of primary tumors at early stage, classified as ``metastatic" and ``non-metastatic" after follow-up. In ``Loi" we considered a total of 125 samples, 28 metastatic and 97 non-metastatic. In ``Frid" we considered a total of 117 samples, 26 metastatic and 91 non-metastatic.\\
Since the microarray platform was the same for the three datasets, the intersection with the PPI network resulted in the same network of about 5000 nodes, with a giant component of 4559 nodes considered for the analysis.\\
For each dataset, a restricted gene list was obtained by performing a Student's T test for uncoupled samples over the two different groups (relapsing vs. not relapsing, metastatic vs. non-metastatic) in order to look for genes significantly changing expression value. With a $P=0.05$ significance threshold, for the ``Wang" dataset we obtained a sub-network of 693 genes that, after the isolated node removal, was reduced to 455 genes. For the ``Loi" dataset we obtained a sub-network of 384 genes, reduced to 212 genes. For the ``Frid" dataset we obtained a set of 195 genes, that were reduced to 85 nodes with at least one interaction. No post-hoc correction for multiple test was applied, in order to keep a sufficient number of significant genes (after Benjamini-Hochberg correction 0 probes would have left in ``Loi" and ``Frid" datasets, and only 47 in ``Wang" dataset).

\subsubsection*{Ageing dataset}
We considered a cross-sectional study (time series) of 25 whole-genome expression profiles of T lymphocytes extracted from healthy males of ages spanning typical adult human lifespan (from 25 to 97 years, see \cite{Remondini} for further details).\\ 
This dataset could be divided into 5 age groups with about 10 y between each group: A) 25-34 y (mean $=$ 29.6 y); B) 43-46 y (mean $=$ 44 y); C) 55-62 y (mean $=$ 58.2 y); D) 70-79 y (mean $=$ 74.2 y); E) 92-97 y (mean $=$ 94.4 y).\\
The gene expression dataset (obtained through a custom array, see \cite{Remondini}) after processing is composed by 13103 probes x 25 age samples. The intersection with the giant component of Pathway Commons data gives a network of 6374 nodes, with a giant component of 6353 nodes considered for analysis.\\
A restricted gene list was obtained by performing a 1-way Anova over the age groups, in order to look for genes significantly changing expression profile in time. A total of 1600 probes were found to be differentially expressed at 5$\%$ significance level after Benjamini-Hochberg post-hoc correction. In this case the resulting PPI network used in our analysis consisted of 638 nodes.\\

\subsection*{Entropy of network ensembles $S_{NE}$ and its application to biological networks}
The main idea of this approach \cite{Bianconi,BianconiPNAS} is that every real network (the PPI network in our case) can be seen as a specific instance of a general ensemble of networks that satisfy the following constraints:
\begin{itemize}
\item Fixed number of interactions per each node (given by PPI network degree sequence $\{k_i\}$ )
\item Fixed number of interactions per each bin obtained from the empirical distribution of the distances $\{ d_{ij}\}$ between the expression level of genes $i$ and $j$, depending on the chosen metrics (e.g. Pearson's Correlation coefficient $R$, $R^2$ or Euclidean distance)
\end{itemize}

We consider a canonical approach (in which the constraints are satisfied on average by the network ensemble) but also a microcanonical approach could be possible (in which each network of the ensemble satisfies exactly the constraints). The two approaches are related by the entropy for large deviations $\Omega$ \cite{Bianconi}.\\ 

We define the canonical entropy of a network ensemble as
\begin{equation}
S_{NE}=- \sum_{i<j}p_{ij}\log p_{ij} - \sum_{i<j}(1-p_{ij})\log (1-p_{ij})
\end{equation}
The goal is the computation of $p_{ij}$ (probability of having a link between every $i$-th and  $j$-th node) given the appropriate constraints stated as functions of the probability matrix $\{ p_{ij}\}$.\\
We define the \textit{spatial ensemble}  imposing the constraints of given degree sequence $\{ k_i \}$ and number of links at distance $d\in I_l$, $B_l$), described by the following equations:

\begin{eqnarray}
\label{constraint1}
k_i=\sum_j^N p_{ij}; \qquad i=1, ..., N\\
\label{constraint2}
B_l=\sum_{i<j}^N\chi_l(d_{ij})p_{ij}; \qquad l = 1, ..., Nb
\end{eqnarray}
where $N$ is the number of nodes in the network, $Nb$ is the number of bins considered for the empirical distribution of distances, and $\chi_l$ is the characteristic function of each bin of width $(\Delta d)_l$: $\chi_l(x) = 1$ if $x \in [d_l,d_l+(\Delta d)_l]$, $\chi_l(x) = 0$ otherwise.
The value of canonical entropy for a spatial ensemble is obtained by maximization of the following equation:
\begin{equation}
\frac{\partial}{\partial{p_{ij}}} \left \{ S_{NE} + \sum_i^N \lambda_i \left( k_i -\sum_j p_{ij} \right)+\sum_l^{Nb}g_l\left(B_l-\sum_{i<j}^N\chi_l p_{ij} \right)
  \right \}=0
\end{equation}
where ${\lambda_{i}}$ and ${g_l}$ are the the Lagrangian multipliers related to our constraints. 
For each $(i,j)$ the resulting marginal probability is
\begin{equation}
\label{prob_ensemble}
p_{ij}=\sum_{l}^{Nb}\chi_{l}(d_{ij})\frac{e^{-(\lambda_{i}+\lambda_{j}+g_{l})}}{1+e^{-(\lambda_{i}+\lambda_{j}+g_{l})}}=\sum_{l}^{Nb}\chi_{l}(d_{ij})\frac{z_{i}z_{j}W_{l}}{1+z_{i}z_{j}W_{l}}
\end{equation}
where $z_i=e^{-\lambda_{i}}$, $W_{l}=e^{-g_l}$, commonly known as hidden variables,  are functions of the Lagrangian multipliers $\lambda_{i}$ and $g_l$.
The $W_l$ vector distinguishes the \textit{spatial ensemble}  from the \textit{configuration ensemble} approach, in which only the degree sequence constraints are given (same equation as before but only with the constraints as in Eq. \ref{constraint1}).
The vector $W_l$ contains the information about gene expression profiles, modulating the probability $p_{ij}$ of having a link between the node $i$ and the node $j$ with a given expression difference $d_{ij}$.
A significant difference between the canonical entropy calculated in the spatial and configuration ensembles reflects the relevance of the information encoded in the distance matrix, and thus in the integration of gene expression data onto the PPI network (see the Supplementary Material for an explicative example).\\
The network entropy $S_{NE}$ is related to the number of possible networks that satisfy the chosen constraints, that in our case are the PPI degree sequence and the distribution of distances between nodes (number of links per bin). Since in our case for each dataset the degree sequence is fixed, the entropy value for each sample is characterized by its gene expression profile.
A high value of $S_{NE}$ means that a large number of networks satisfy the given constraints, thus the set of features associated to the observed network (given by the gene expression values) is not very particular (i.e. informative from an information theory point of view \cite{Cover}).
We can say that there is a high degree of \textit{plasticity} for the system in this case, since a wide range of parameter values (gene expression profiles) are allowed to the cell. On the contrary, if the network has lower entropy, only few configurations can satisfy the given constraints, and the cell is in a more specific state with less degrees of freedom. This can be interpreted as 1) the system has been more finely tuned or 2) the system has a smaller range available to its parameters. We remark that a positive or negative interpretation of the entropy values must be related to the context in which the measure is performed, as it will be shown below in our analyses.\\
Gene expression data are used to construct the distance matrix $d_{ij}$: the choice of a suitable metrics can improve the information content encoded in it. We chose an euclidean metrics (see below) instead of a correlation-based measure as in \cite{Severini} since it allowed the calculation of an entropy value for each sample, useful for the performed statistical analyses.
Since the statistical weight $W_l$ of each bin $l$ is related to the number of interactions that fall within it, also the choice of distance binning is important. Even if the entropy values depend on the number of bins (whose choice is a tradeoff between an accurate representation of the empirical distribution and an adequate number of samples falling into each bin) the achieved results should be robust to small perturbations of this number. We performed our analysis with linearly spaced bins, considering a range of bin numbers depending on the size of our network: approximately $70\pm 10$ bins for networks with more than 1000 nodes (the whole-genome networks), $25\pm 5$ bins for smaller networks (the KEGG-induced subnetworks and the statistically significant gene subsets).\\ 
For the calculation of the entropy values and the Lagrangian multipliers, we developed an iterative algorithm: given a strarting guess for the value of the lagrangian multipliers $\{z_i\}$ and $\{W_l\}$, the $p_{ij}$ values are calculated according to Eq. (\ref{prob_ensemble}). These values are then subsituted in the constraint equations (\ref{constraint1},\ref{constraint2}) for the next calculation of the lagrangian multipliers, and the process is repeated upon convergence.
We checked by random sampling that the application of the iterative algorithm for different initial guesses would lead to the same entropy values.
In our analyses, the algorithm convergence threshold was set to $10^{-4}$, and we remark that every significant change in entropy values was at least of a order of magnitude higher.
This algorithm is available both in Matlab and Python code. 

\subsubsection*{Multi-scale entropy approach}
Taking advantage of the a priori biological knowledge available from the KEGG database (http://www.genome.jp/kegg/pathway.html) it is possible to obtain several induced subnetworks on the PPI network: the genes annotated in the PPI can be divided into 6 functional groups (``Metabolism", ``Genetic Information Processing", ``Environmental Information Processing", ``Cellular Processes", ``Organismal Systems" and ``Human Diseases"), that can be further subdivided into 42 metapathways and again into about 260 single biological pathways (see Supplementary Table 3).
We found 1976 non-isolated nodes for the ageing dataset and about 1530 for the cancer datasets annotated both in PathwayCommons and KEGG: starting from these genes we calculated $S_{NE}$ for the subnetworks induced by the 6 functional groups, populated by several hundred genes each (see  Tables \ref{tabKEGGPathCancerWang}, \ref{tabKEGGPathCancerLoi}, \ref{tabKEGGPathCancerFrid}, \ref{tabKEGGPathCancerWangSel}, \ref{tabKEGGPathCancerLoiSel}, \ref{tabKEGGPath} ). 
We did not scale the analysis down to the single KEGG metapathways and pathways, because most of the resulting PPI subnetworks were very small ($N<30$ nodes) and we were not confident that the subnetwork features could be considered relevant, but it is not excluded that other larger datasets could be scaled down to such level.

\subsubsection*{Cancer dataset analysis}
In order to evaluate the differences between gene expression values for each specific sample, we chose the Euclidean distance as a metrics: for each sample $a$ (with $a$ ranging according to dimension of the considered dataset), we define the distance $d_{ij}^a$ between $i$-th gene expression value $g_i^a$ and $j$-th gene expression value $g_j^a$ as
\begin{equation}
d_{ij}^a=\sqrt{(g_i^a-g_j^a)^2}=|g_i^a-g_j^a|
\label{euclideandist}
\end{equation}
In this way we obtain 286 $S_{NE}$ values for ``Wang" (107 relapsing and 179 not relapsing), 125 values for ``Loi" (28 metastatic and 97 non-metastatic), 117 values for ``Frid" (26 metastatic and 91 non-metastatic).\\
We have the same number of samples both for the whole dataset and its 5$\%$ significance level selection. This allowed us to perform significance tests (Student's T test) between the pooled entropy values of relapsing vs. not relapsing and metastatic vs non-metastatic groups.

\subsubsection*{Ageing dataset analysis}
Ageing mechanism can be considered as a really different cellular perturbation compared to cancer and, moreover, the studied ageing dataset has a specific time sequence (given by samples' age, since it is not an ``old vs. young" experimental design) that encodes relevant information about the mechanism to be studied. 
In order to evaluate the differences among gene expression values at a specific age, we chose the Euclidean distance also for this dataset as in Eq. \ref{euclideandist}.
In this way we obtained 25 $S_{NE}$ values, one for each sample, both for the whole dataset or the 5$\%$ significance level selection, that allow us to perform significance tests between the age groups (by 1-way ANOVA or multiple Student's T tests between age groups).

\section{Results/Discussion}

\subsection*{Cancer dataset}

Considering the entropy values calculated over the whole-genome datasets, the ``Loi" dataset showed a significant difference between the two groups with Student's T Test ($P=0.023$). The mean entropy value for non-metastatic group was higher than the metastatic  ($S_{NE}^{non-met}=61.71$, $S_{NE}^{met}=61.70$). The same trend was observd in ``Wang" and ``Frid" datasets, even if not significant at the whole-genome level (see Tab. \ref{table1}) .\\
Taking the gene selection with 5$\%$ significance level (single-gene T test between groups for each dataset) we found significant differences in entropy values for both ``Wang" ($P=4.3\cdot 10^{-5}$) and ``Loi" ($P=0.0012$) datasets, with the same trend observed at whole-genome level, that is, with a greater mean value for the entropy of not relapsing and non-metastatic samples (see Tab. \ref{table2}).\\
We remark that the samples of all datasets were taken at the onset of primary tumor, and the ``relapsing" and ``metastatic" labels were assigned after a follow-up analysis: thus we are not performing a direct comparison between metastatic and primary tumour tissues, but they are all primary cells that have eventually progressed into metastatic or relapsing.
In the light of this remark, a higher entropy value for the primary tumour state in both datasets suggests that the cells that will become relapsing or metastatic have a more controlled expression profile, as if for example a large group of genes could undergo (in a coordinated way) only strong activation or inactivation.
We interpret this result as the occurrence of an initial ``tuning" of expression profiles in order to achieve the following cellular state (metastatic/relapsing), for example by silencing or overexpressing specific genes that should otherwise be less tightly regulated \cite{HallmarkCancer00}.
This result does not exclude that, after the primary tumour has evolved into a metastatic or relapsing phenotype, the cell moves towards a more de-differentiated/deregulated state (as for metastatic cells in most cases \cite{dediff1,dediff2,dediff3}).

\begin{table}[!t]\small
\caption{Values of $S_{NE}$ for the whole-genome cancer datasets, with the P value obtained by Student's T test.\label{table1}}
{\begin{tabular}{llll}\toprule
\quad & Rel/Met & Not Rel/Not Met & P-value\\
Wang & 61.6704 & 61.6734 & 0.3194\\
Loi & 61.7031 & 61.7134 & 0.0233 \\
Frid & 61.3098 & 61.310 & 0.9631\\\botrule
\end{tabular}}{}
\end{table}

\begin{table}[!t]\small
\caption{Values of $S_{NE}$ for the $5\%$  selection subset of cancer datasets, with the P value obtained by Student's T test.\label{table2}}
{\begin{tabular}{lllll}\toprule
\quad & Rel/Met & Not Rel/Not Met & Size & P-value\\
Wang & 18.7716 & 18.7807 & 455 & 4.25 $\cdot 10^{-5}$\\
Loi & 13.0773 & 13.1059 & 212 & 0.0012 \\
Frid & 6.2603 & 6.2590 & 85 & 0.6898 \\\botrule
\end{tabular}}{}
\end{table}

\subsubsection*{Multi-scale analysis of the cancer dataset}
We computed the entropy $S_{NE}$ for each sample also in different PPI subnetworks induced by the functional classifications of genes in the KEGG database; $S_{NE}$ was evaluated both at a whole-genome level and on a $5\%$ significance threshold gene selection (see Tab. \ref{tabKEGGPathCancerWang}, \ref{tabKEGGPathCancerLoi}, \ref{tabKEGGPathCancerFrid}, \ref{tabKEGGPathCancerWangSel},  \ref{tabKEGGPathCancerLoiSel}). The results for the $5 \%$ selection of Frid dataset always produced networks with $N\leq10$ nodes, so we excluded them from the analysis.\\
The PPI network was divided into 6 six biological functional groups (see Methods Section). 
The ``Metabolism" group was significant for the ``Wang" dataset only at whole-genome level ($P=0.039$) with a higher entropy value for the non relapsing case, while the ``Genetic Information Processing" group was significantly different both at a whole-genome level ($P=0.0044$) and in the $5\%$ selection ($P=0.00011$), with higher entropy values for the relapsing samples.
The ``Environmental Information Processing" group was significantly different in the $5\%$ selection for ``Wang" ($P=0.035$) and ``Loi" ($P=0.028$) datasets, with a higher entropy value in the relapsing/metastatic cases.
The increase in entropy in the most aggressive cases, related to the the ``Information Processing" functional groups, might reflect a larger deregulation, or a wider parameter landscape that can be explored by the cells in such a state, as compared to less aggressive phenotypes (that did not relapse nor developed metastases).

\begin{table}[!t]\small
\caption{Wang cancer data set: functional groups for the whole-genome cancer dataset (entropy values, network sizes and P-value of the Student's T test). \label{tabKEGGPathCancerWang}} 
{\begin{tabular}{lllll}\toprule
KEGG group & Rel & Not Rel & Size & P value \\
Metabolism &	12.5810 & 12.5838 & 364 & 0.0386 \\
Genetic Information Processing &	21.6013  & 21.5540  &  143  &  0.0044 \\
Environmental Information Processing & 14.4845 & 14.4848 & 658 & 0.7834 \\
Cellular Processes &	19.3270 & 19.3262 & 303 & 0.4733 \\
Organismal Systems &	17.0946 & 17.0940 & 491 & 0.6829 \\
Human Diseases & 11.9827 & 11.9834 & 238 & 0.7157 \\\botrule
\end{tabular}}{}
\end{table}

\begin{table}[!t]\small
\caption{Loi cancer data set: functional groups for the whole-genome cancer dataset (entropy values, network sizes and P-value of the Student's T test). \label{tabKEGGPathCancerLoi}} 
{\begin{tabular}{lllll}\toprule
KEGG group &  Met &  Not Met & Size & P value \\
Metabolism &	 12.5735 &    12.5735  &  364  &  0.9996  \\
Genetic Information Processing & 21.5895  &  21.5960 &  143  &  0.7510 \\
Environmental Information Processing & 14.4759 & 14.4743 & 658 & 0.1724 \\
Cellular Processes & 19.3337 & 19.3304 & 303 & 0.0344 \\
Organismal Systems & 17.0808 & 17.0862 & 491 & 0.0228 \\
Human Diseases & 11.9942 & 11.9964 & 238 & 0.3439 \\\botrule
\end{tabular}}{}
\end{table}

\begin{table}[!t]\small
\caption{Frid cancer data set: functional groups for the whole-genome cancer dataset (entropy values, network sizes and P-value of the Student's T test).\label{tabKEGGPathCancerFrid}} 
{\begin{tabular}{lllll}\toprule
KEGG group & Met & Not Met &	Size & P value \\
Metabolism & 12.3057  &   12.3035  &  363  &  0.2498 \\
Genetic Information Processing & 21.6723  &  21.6422  &  143 & 0.2283 \\
Environmental Information Processing & 14.3175 & 14.3168 & 657 & 0.6659 \\
Cellular Processes & 19.0730 & 19.0729 & 302 & 0.9632 \\
Organismal Systems & 16.7771 & 16.7747 & 489 & 0.2078 \\
Human Diseases & 11.3462 & 11.3472 & 235 & 0.7222 \\\botrule
\end{tabular}}{}
\end{table}

\begin{table}[!t]\small
\caption{Wang cancer data set: functional groups for the $5\%$  selection subset (entropy values, network sizes and P-value of the Student's T test).\label{tabKEGGPathCancerWangSel}} 
{\begin{tabular}{lllll}\toprule
KEGG group & Rel & Not Rel & Size & P value \\
Metabolism &	3.3180 & 3.3153 & 46 & 0.4841 \\
Genetic Information Processing &	6.2894  & 6.2301  &  33  &  0.0001 \\
Environmental Information Processing & 3.9513 & 3.9447 & 69 & 0.0351 \\
Cellular Processes &	3.9491 & 3.9658 & 37 & 0.0096 \\
Organismal Systems &	3.6396 & 3.6422 & 55 & 0.5510 \\
Human Diseases & 2.0865 & 2.0865 & 36 & 0.9982 \\\botrule
\end{tabular}}{}
\end{table}

\begin{table}[!t]\small
\caption{Loi cancer data set: functional groups for the $5\%$  selection subset (entropy values, network sizes and P-value of the Student's T test).\label{tabKEGGPathCancerLoiSel}} 
{\begin{tabular}{lllll}\toprule
KEGG group &  Met &  Not Met & Size & P value \\
Metabolism & 3.1166 & 3.1066  & 27  &  0.5269  \\
Genetic Information Processing & 1.815  &  1.7987 &  10  &  0.6001 \\
Environmental Information Processing & 1.8550 & 1.8380 & 28 & 0.0284 \\
Cellular Processes & 3.6077 & 3.6047 & 26 & 0.7531\\
Organismal Systems & 3.0037 & 2.9901 & 33 & 0.0096 \\
Human Diseases & 1.9222 & 1.9054 & 18 & 0.2091\\\botrule
\end{tabular}}{}
\end{table}

\subsection*{Ageing dataset}

\begin{figure}[!tpb]
\includegraphics[width=\textwidth]{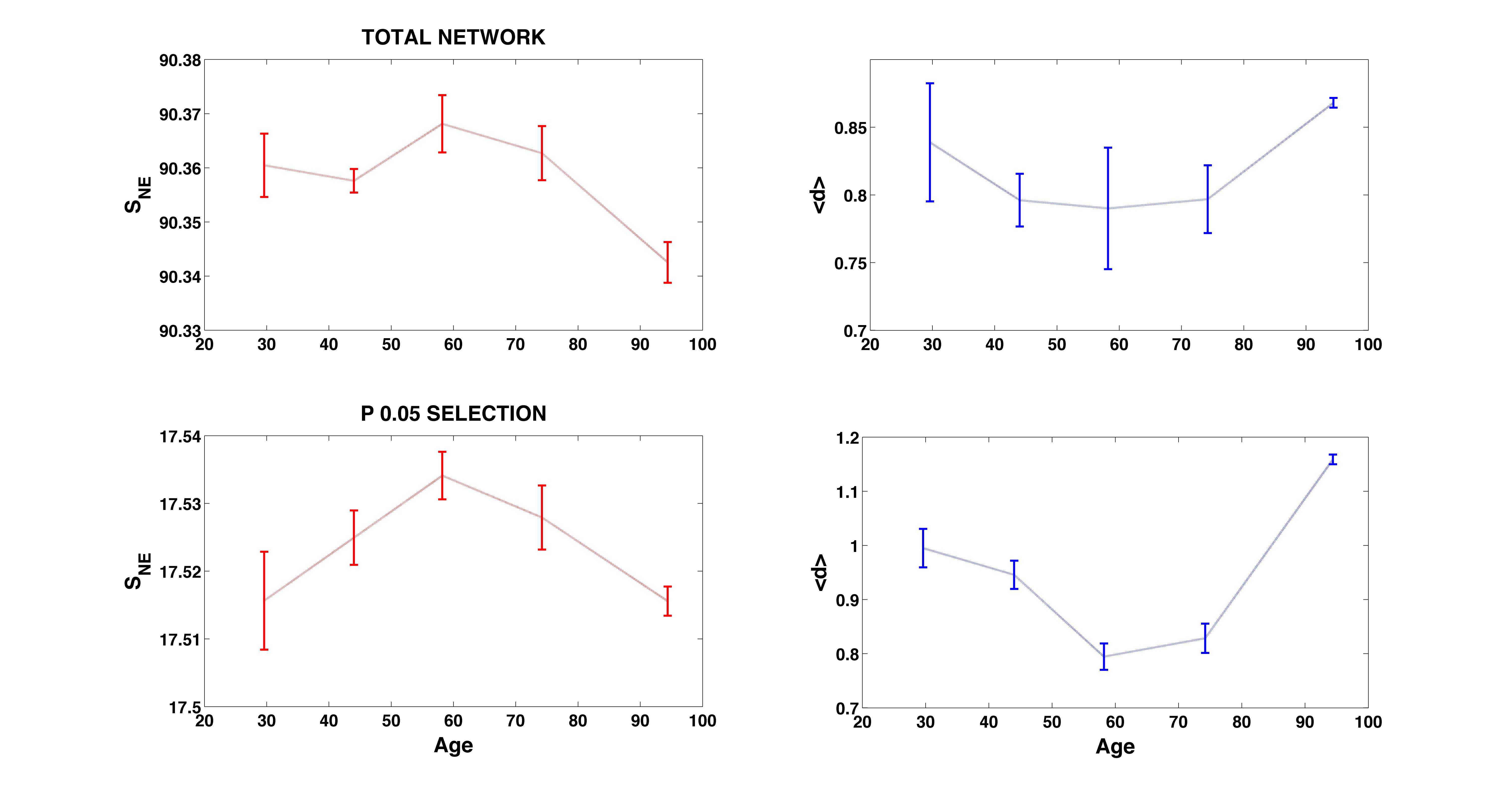}
\caption{Error plots for the Ageing dataset. 
Left boxes: plot of $S_{NE}$ over the 5 age groups, for the whole-genome PPI network (top) and for the $5\%$ significance level selection (bottom).  
Right boxes: error plots of the mean values of all the euclidean distances, at a whole-genome level (top) and for the selection (bottom).}\label{PAPERAgeingEuclidean}
\end{figure}

The entropy values as a function of age are presented in Fig. \ref{PAPERAgeingEuclidean}. For both the whole-genome network and the $5\%$ significance level selection network, we plotted the mean value of $S_{NE}$ and its standard deviation of the mean (SDOM) for each age group. 
For the whole dataset, we observe the smallest values of $S_{NE}$ in the fifth group, the so-called ``successfully aged" people SA (that live longer than average life expectancy of about 80 years, see for example www.indexmundi.com) that behaves in a significantly different way from the rest of the age groups: we have $P=0.011$ by 1-way ANOVA, with significant differences between group 3 and group 5 ($P=0.0052$) and group 4 and group 5 ($P=0.013$) calculated by Student's T test as a post-hoc test. In the same figure, we plot the time trend of the average distance for each sample, that shows no direct relationship with the entropy values, confirming that they represent a different kind of information.
This decreasing trend in entropy means that the number of possible gene expression profiles that lead to a SA phenotype are lower, in agreement with the fact that only a small subset of total population will become SA, and also that since the possible profiles allowed in this state are less, SA subjects may have a reduced range of responses to stimuli and perturbations, implying a minor plasticity.\\

For the $5\%$  selection subset we observe a non monotonic trend in time: the smallest mean values are observed for the youngest and the SA group, while in the middle-aged groups entropy increases and reaches its maximum in the third age group. Moreover, the SA individuals have the smallest standard deviation and they are significantly different with the third age group  ($P=0.043$ by 1-way ANOVA over the five age groups, and $P=0.0032$ between group 3 and group 5 with post-hoc Student's T test). This result supports the fact that SA individuals maintain specific gene expression profiles unchanged in time (as compared to the young group) while the middle and old age groups (that have a very low probability to become SA) show a significant drift from this range of `optimal'' values. We remark that this definition of optimality assumes that healthy young samples represent the best case of cell performance, and that $S_{NE}$ is not a simple function of gene expression values, but integrates the differences between expression levels of genes with the network of proteins that are supposed to interact, thus taking into account the mutual relationships between them.
The increase in entropy in the middle groups suggest an increasing global deregulation or loosening of the expression ranges with age, that is kept controlled in the SA samples.\\
The relevance of the information encoded in gene expression values for the calculation of entropy is reflected by the fact that configuration entropy values (that considers only the constraints on PPI network degree sequence, as described in the Methods Section) are significantly higher than $S_{NE}$ values, for both the whole-genome and the selection dataset. For the whole-genome dataset the average value of spatial entropy over the 25 samples is $<S_{NE}>=90.358\pm 0.013$ (1-$\sigma$ confidence interval) and the configuration entropy $S^{conf}_{T}=90.412$ (equal for all the samples since it depends only on the PPI), so their difference is about 4.14 $\sigma$; for the selection dataset we have $<S_{NE}>=17.524\pm 0.012$ and $S^{conf}_{SEL}=17.573$, so $\Delta S = 4.13$ $\sigma$.
We remark that spatial and configuration entropy values become more and more similar when the distance distribution approaches a random uniform distribution, thus when no information is encoded in the distance matrix (see Supplementary Material Section 2).\\

\subsubsection*{Multi-scale analysis of the ageing dataset}

\begin{figure}[!tpb]
\centerline{\includegraphics[width=\textwidth]{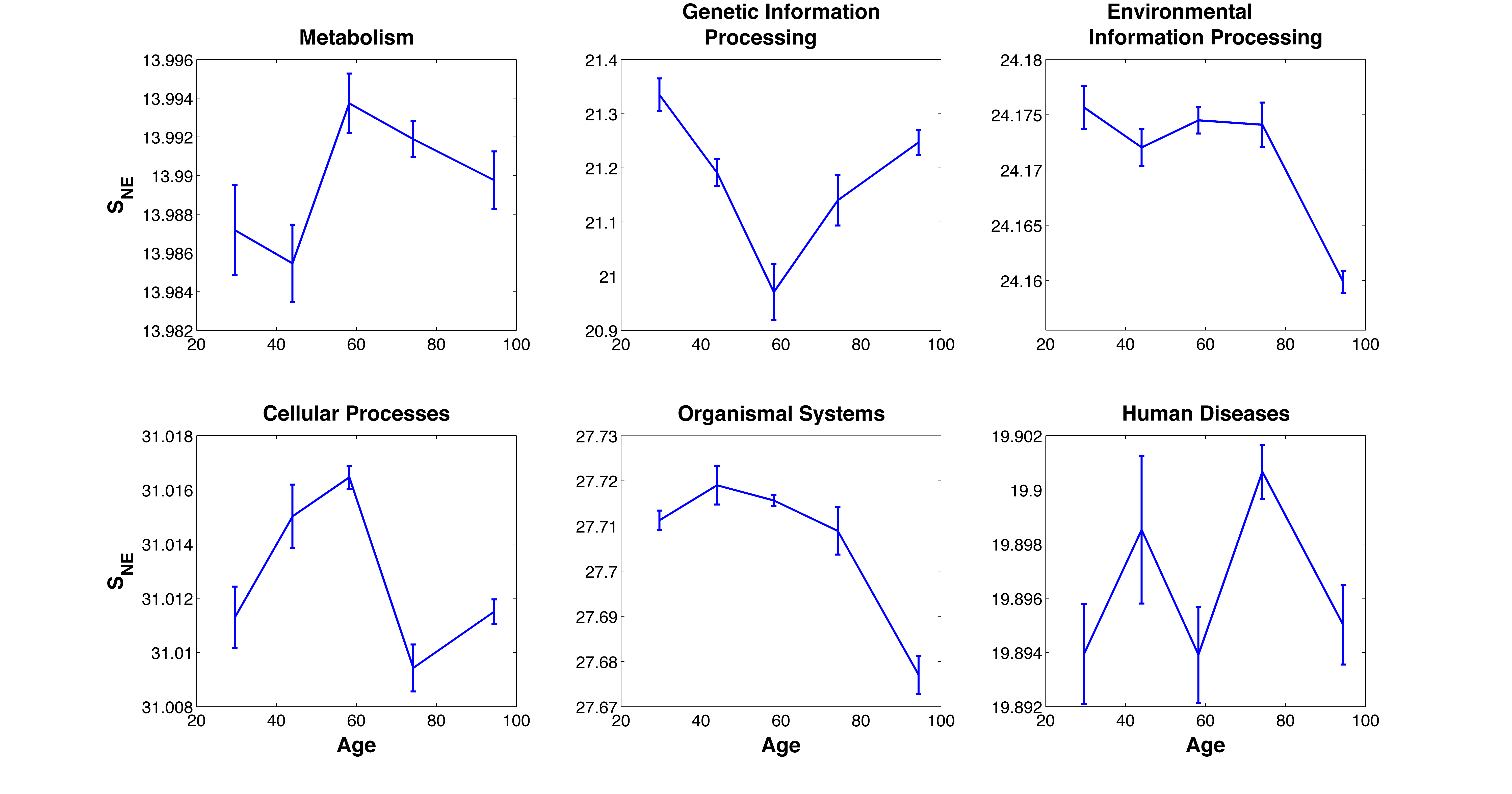}}
\caption{Entropy trends of the 6 functional groups groups for the ageing dataset. We underline the similar decreasing trend for both ``Environmental Information Processing" and ``Organismal Systems". We also spot a significant non-monotonic behavior of ``Genetic Information Processing".}\label{Path}
\end{figure}

We computed the entropy $S_{NE}$ for each sample also in 6 PPI subnetworks induced by the functional groups as in the KEGG database. 
Considering at least one significant difference between two age groups (evaluated by Student's T test, see Fig. \ref{Path} and Tab. \ref{tabKEGGPath}) we observed both monotonic and non-monotonic trends.
We interpreted the monotonic trends as markers of a progressive ageing of the cell, whereas the non-monotonic trends could be associated to successful ageing, characterizing functionalities that are preserved from the young to the old age.
For the whole-genome dataset,  the ``Environmental Information Processing" and the ``Organismal Systems" groups showed a significant decrease in entropy mainly for the SA group, that might represent a reduction of \textit{plasticity} in terms of the possible expression profiles achievable by the cell. This could be associated to the so-called ``frail" phenotype \cite{frailty}, a state of high sensitivity and inability to respond to perturbations (e.g. traumas, illnesses) that leads to irreversible damage of organismal functionality.
We remark that our dataset consisted of healthy subjects, thus not classified as frail with the existing definitions \cite{frailty_index}: it could be interesting to verify if the expression profiles of frail samples would show an early decrease of entropy as compared to non-frail subjects of the same age, enforcing our hypothesis of frailty as a state of reduced plasticity, and suggesting entropy as a possible marker of this pathologic state.
Interestingly, the ``Metabolism" group has an opposite trend in which entropy increases in the elderly (even if it is not statistically significant), as if a more strict regulation necessary for the metabolic processes would be loosening with age.
We remark that the same trend for $S_{NE}$ has a different meaning in relation to the biological context: it is plausible that metabolic processes have to be strictly regulated in order to achieve an optimal cellular functionality, while signalling processes must be able to respond to different internal and environmental stimuli, thus keeping a wide range of possible responses.\\
The ``Genetic Information Processing" and the ``Cellular Processes" functional groups show a non monotonic trend, with opposite behavior in the median groups. The ``Genetic information processing" functional group shows a decrease in entropy for people of median age, that is in agreement with the decline in plasticity observed in the ``Environmental Information Processing" group, whereas the ``Cellular Processes" group (related to catabolism, cell cycle and transport) is in agreement with the interpretation of the trend for the ``Metabolism" group.
No clear trend was observed for the ``Human Diseases" group, and no significant differences were found between age groups in this case.
For the $5\%$ significance subset, we observed a good agreement of the trends with respect to the global dataset, even if the number of nodes in the subnetworks is very low (e.g. an increase in ``Metabolism", a non-monotonic trend for ``Genetic Information Processing" and ``Cellular Processes" groups, see Supplementary Materials Section 1).

The results for the ageing dataset are in qualitative agreement with previous analyses \cite{Remondini}, showing also non-monotonic trends for the same dataset, even if we have shown that there is not a trivial relation between entropy and gene expression values. 

The same analysis performed with the entropy-like measure described in \cite{Severini} showed a strong dependence on the number of nodes in the network, with a systematic reduction of statistical significance for the $5\%$ subsets, both for the cancer and the ageing datasets (as shown in Supplementary Materials Section 4).

\begin{table*}[!t]\footnotesize
\caption{Functional groups for the whole-genome ageing dataset: network sizes and significant comparisons (Student's T test). Note: italics values are weakly non significant ($P\simeq 0.07$). \label{tabKEGGPath}}
{\begin{tabular}{llll}\toprule
KEGG group & Size & P value (ANOVA) &Post-hoc test P value (Groups) \\
Metabolism & 488 &  0.5615&\\
Genetic Information Processing & 177 & \textit{0.0612} &0.032 (1-3),   \textit{0.075} (3-5)\\
Environmental Information Processing & 849 & 0.0340&0.018 (1-5), 0.029 (2-5), 0.0032 (3-5),   0.031 (4-5)\\
Cellular Processes & 465 & 0.1040 & 0.018 (3-4),   0.0071 (3-5)\\ 
Organismal Systems & 611 &  0.0141& 0.018 (1-5),   0.014 (2-5),   0.012 (3-5), \textit{0.068} (4-5)\\
Human Diseases & 317 & 0.7087&\\\botrule
\end{tabular}}{}
\end{table*}


\section{Conclusion}
We have applied an entropy measure based on a rigorous statistical mechanics definition, that integrates the protein interaction network with gene expression profiling, to characterize different levels of cellular perturbation (cancer subtypes and physiological ageing).
This measure characterizes the number of networks that satisfy given constraints (namely the PPI network degree sequence and the empirical distribution of distances between gene expression profiles), and can be interpreted as the extent of the ``parameter space" allowed to the cell in a given state. 
We showed that, due to the imposed constraints, it is not just a measure of cellular ``disorder": depending on the considered biological function, a high value of entropy can represent a high degree of cellular plasticity, seen as the ability to adjust its internal state to the incoming stimuli (thus with a positive connotation) or a high level of deregulation in terms of the range of values allowed to the expressed genes (in this case with a negative connotation).\\

These results show different perspectives on the analysis of gene expression data, providing a global view that takes into account single gene measurements and their functional relationships inside the cell mechanisms.
The entropy measure $S_{NE}$ seems an observable sensitive enough to evaluate the effect of physiological perturbations such as the changes occurring during the cellular ageing process, and also the differences between cancer subtypes before the progression to metastatic and relapsing phenotypes. 
The statistical significance of $S_{NE}$ resulted independent on network properties, such as the number of nodes, and increased when a selected subset was considered, thus reflecting the biological relevance of the data used.

We were able to scale the analysis at different levels based on a priori biological knowledge (as obtained from KEGG database) dividing the full PPI network into subnetworks related to specific biological functions and pathways. The number and type of subnetworks depend on the a priori biological knowledge available (e.g. for the organism studied) and on the number of nodes of the initial PPI network (in order not to obtain too small subnetworks, even if this point needs further study).\\

For the cancer datasets, in the full PPI case we observed a decrease in entropy as compared to the primary tumours (both in the whole-genome and in the selection datasets), that may reflect a more refined tuning of the expression levels in order to reach the metastatic/relapsing states.\\
For the ageing dataset, we observed different trends over time.
Two groups, related to ``Environmental Information Processing" and ``Organismal Systems", showed a monotonic decrease in entropy: we can interpret it as if the parameter space explored by the cells is reducing, allowing less plasticity in cell response.
The ``Metabolism" group had an increasing trend in entropy, that could be associated to a progressive deregulation of its functionality.
The ``Genetic Information Processing" and ``Cellular Processes" groups showed a non-monotonic trend, in which the oldest and the youngest samples have similar values of entropy, thus they might represent  a set of functionalities conserved through lifetime in the successfully aged people.\\

\section*{Acknowledgement}
DR and GC were supported by EU project MIMOmics n. 305280 and FibeBiotics n. 289517, and TO61 INFN initiative.

\bibliographystyle{plain}
\bibliography{network_entropy_bib}

\newpage

\section*{Supplementary Material}

\subsection*{Analysis of the functional groups for the 5 $\%$ selection of the ageing dataset}

\begin{figure}[h!]
\label{agesel_fig}
\centering
\includegraphics[width=150 mm]{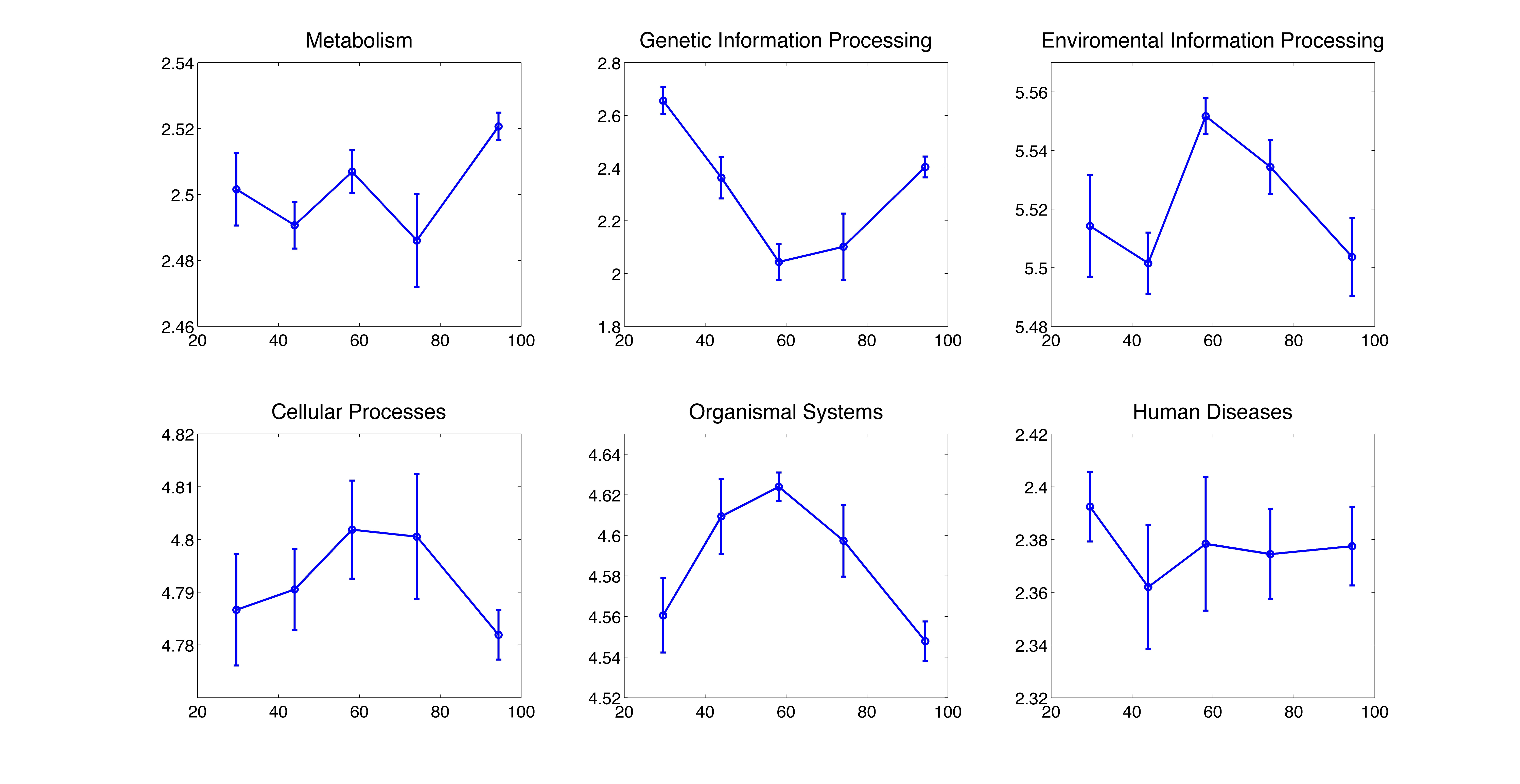} 
\caption{Plot of the $S_{NE}$ trends}
\end{figure}

In this section we show the results for the 5 $\%$ selection of the ageing dataset. 
In Fig. \ref{agesel_fig} we can observe which trends are qualitatively similar to those of the whole-genome dataset: same increasing trend for the "Metabolism" functional group (with a significant difference between two age groups, see Tab. \ref{agesel_tab}); same non-monotonic trend for the "Genetic Information Processing" functional group, with significant differences between middle age group and extremal age groups; same non-monotonic trend for the "Cellular Processes" functional group. We remark that the "Environmental Processing" and "Organismal Systems" functional groups differ in the first part of the trend between the whole-genome and the $5\%$ selection datasets, but they are similar to the respective trends of the the whole PPI network $S_{NE}$ values.

\begin{table}[!h]\small
\label{agesel_tab}
\centering
{\begin{tabular}{|l|l|l|l|}\hline
KEGG group & Size & P (ANOVA) &Post-hoc test P value (Groups) \\
\hline
Metabolism & 44 &  0.1073 & P=0.0095 (2-5)\\
\hline
{Genetic Information Processing} &  &  & P=0.0171 (1-2),  P=0.0001 (1-3)\\& 17& 0.0001& P=0.0084 (1-4), P=0.0054 (1-5)\\& & & P=0.0054 (2-3), P=0.0033 (3-5) \\
\hline
{Environmental Information Processing} & 94 & 0.0323 & P=0.0051 (2-3), P=0.0465 (2-4)\\& & & P=0.0179 (3-5) \\
\hline
Cellular Processes & 58 & 0.4816 & \\ 
\hline
{Organismal Systems} & 81 &  0.0082 & P=0.0222 (1-3),   P=0.0253 (2-5)\\& & & P=0.0003 (3-5), P=0.0483 (4-5)\\
\hline
Human Diseases & 34 & 0.8651 & \\
\hline
\end{tabular}}{}
\caption{Functional groups for the 5 $\%$ selection of the ageing dataset: network sizes and significant comparisons (1-way ANOVA and post-hoc Student's T test between age groups).}
\end{table}

\newpage

\subsection*{Characterization of configuration and spatial ensembles: a toy model}
\begin{figure}[h!]
\centering
\includegraphics[width=100 mm]{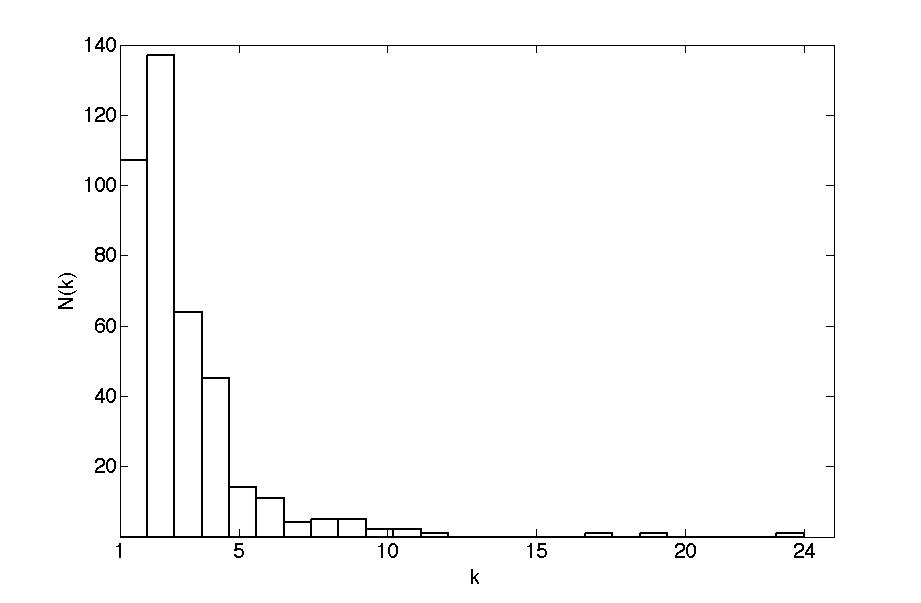}
\caption{Degree distribution of the network: $N=400$ nodes, 558 total links, $k_{min}=1$, $k_{max}=24$, $\left<k\right>=2.8$.}\label{pk}
\end{figure}

We present here a toy model for a better understanding of entropy of network ensemble. We want to underline how the quantity and the quality of the imposed constraints affect the entropy value.\\
We consider only canonical ensembles, i.e. the imposed constraints are satisfied on average. We consider ``simple" networks, namely, in which a link between two nodes $i$ and $j$ can be present ($a_{ij}=1$) or absent ($a_{ij}=0$), no self-interactions are permitted and the matrix $\{ a_{ij} \}$ is symmetric. An ensemble of simple networks is described by the probability matrix $\{ p_{ij} \}$ where, for each $(i,j)$, a link is present with probability $p_{ij}$ or absent with probability $(1-p_{ij})$. Given that matrix elements are independent and uncorrelated random parameters, we obtain a definition of entropy for a network ensemble, called also the Shannon entropy of network ensembles
\begin{equation}
S_{NE}=- \sum_{i<j}  p_{ij}\log p_{ij} + (1-p_{ij})\log (1-p_{ij})
\end{equation}
The main goal is the computation of $\{ p_{ij}\}$ given the appropriate constraints stated as functions of the probability matrix  ($f_k(\{ p_{ij}\})=F_k$ for $k=1, 2, ... M$). Following the principle of maximal entropy, $\{ p_{ij}\}$ is provided by the maximization of $S_{NE}$ subjected to the chosen constraints
\begin{equation}
\frac{\partial}{\partial{p_{ij}}}\left \{ S_{NE} + \sum_k^M \lambda_k \left( F_k - f_k(\{ p_{ij}\} \right) 
  \right \} =0
\end{equation}
Once outlined a specific network, we consider it as a realization of an ensemble of all the possible networks satisfying some given constraints. The entropy value of the considered ensemble changes according to the chosen constraints. 
The computation of entropy gives us a measure about the specificity of our network, or, on the other hand, tells us how much the chosen constraints define an accurate model of the real network.\\
In this paper, for the modelling of biological networks, we mainly referred to two ensembles:
\begin{description}
\item[Configuration ensemble:] given expected number of interactions per each node (PPI network degree sequence $\{k_i\}$ )
\item[Spatial ensemble:]  given expected degree sequence $\{ k_i \}$ and given expected number of interactions per bin $\{ B_l \}$ obtained from the empirical distribution of the distances $\{ d_{ij}\}$ between the expression level of genes $i$ and $j$, depending on the chosen metrics (we always used the Euclidean distance)
\end{description}
We show an example of two spatial ensembles with the same degree sequence but with a different distance distribution of nodes, in order to understand how much the differences in the two distance distributions affect the entropy values. Furthermore, we consider how much the spatial ensemble differs from the configuration ensemble (that has a subset of the constraints expressed by the spatial ensemble, regarding just the topology of the network). 
Even if the entropy value decreases with the addition of constraints (a smaller subset of networks is available) this decrease could be irrelevant when the information encoded in the new constraints is not discriminating.\\
For our example, we consider a network of 400 nodes with a fat-tailed distribution, in resemblance to the PPI used in the paper (see Fig. \ref{pk}). We investigate three different ensembles:
\begin{enumerate}
\item Configuration ensemble (only fixed degree sequence)
\item Spatial ensemble with same degreee sequence and a distance matrix $\{ d_{ij}^a \}$ calculated starting from a random vector of $N=400$ values randomly distributed between 0 and 5
\item Spatial ensemble as before, but the distance matrix $\{ d_{ij}^b \}$ is calculated starting from the previous random vector in which 100 values are randomly distributed between 0 and 1
\end{enumerate}

For both the distance distributions we chose a linear binning of 20 bins, and we considered 50 possible realisations for each case.\\

The imposed constraints are expressed as a function of the probability matrix $\{ p_{ij}\}$: the configuration ensemble considers only the first constraint, while the spatial ensemble considers both:
\begin{eqnarray}
k_i&=&\sum_j^N p_{ij}; \qquad i=1, ..., N=100\\
B_l&=&\sum_{i<j}^N\chi_l(d_{ij})p_{ij}; \qquad l = 1, ..., Nb=20
\end{eqnarray}
where $N$ is the number of nodes in the network, $Nb$ is the number of bins considered for the empirical distribution of distances, and $\chi_l$ is the characteristic function of each bin of width $(\Delta d)_l$: $\chi_l(x) = 1$ if $x \in [d_l,d_l+(\Delta d)_l]$, $\chi_l(x) = 0$ otherwise.
Therefore, the configuration ensemble has $N=400$ constraints and, the spatial ensembles have  $N+Nb=420$ constraints.\\
We obtained value of entropy for the configuration ensemble (corresponding to a single degree sequence) and a distribution of 50 entropy values for each spatial ensemble: the summary statistics are shown in Tab. \ref{table1}\\
\begin{table}[h!]
\center
\begin{tabular}{|c|c|c|c|}
\hline
Ensemble & $S_{NE}$ & $\sigma$ & $\Delta S$\\
\hline
Configuration ensemble $S_{CONF}$ & 7.60 &   &  \\
Spatial ensemble $\{ d_{ij}^a \}$ $<S_{NE}^a>$ & 7.57 & 0.01 & $1.9\sigma$\\
Spatial ensemble $\{ d_{ij}^b \}$ $<S_{NE}^b>$ & 7.53 & 0.017 & $4.7\sigma$\\
\hline
\end{tabular}
\label{table1}
\caption{Entropy values for the configuration entropy and the two ensembles. 
The difference $\Delta S$ is computed with respect to $S_{CONF}$ rescaled with the average of the two standard deviations.}
\end{table}\\
The small differences in the entropy values are determined by the larger number of constraints associated to the topology of the network (degree sequence, 400 constraints) as compared to the constraints associated to the distance bins (20 constraints). 
The addition of the constraints on distance distribution reduces $S_{NE}$, and with the same number of imposed constraints $S_{NE}$ is lower when the added information is less random.
The differences between the two spatial ensembles is highly significant ($\Delta S = 2.8 \sigma$, average standard deviation).

\newpage
\subsection*{KEGG database hierarchy}

In this Section we present the subdivision of functional groups into metapathways. The further subdivision into about 260 pathways is not shown. We remark that in our analyses we scaled our results to the functional groups level only, because most of the metapathways had a very small number of nodes in each subnetwork ($N<30$).

\begin{table}[h!]\footnotesize
\caption{Table of KEGG Functional groups and metapathways} 
\centering
\begin{tabular}{ |l|l| }
\hline
KEGG Functional Group & Metapathway  \\
\hline
{Metabolism} & Carbohydrate metabolism \\ & Energy metabolism, Lipid metabolism \\ & Nucleotide metabolism  \\ & Amino acid metabolism  \\ & Metabolism of other aminoacids \\ & Glycan biosynthesis and metabolism \\ & Metabolism of cofactors and vitamins \\ & Metabolism of terpenoids and polyketides \\ & Biosynthesis of other secondary metabolites \\  & Xenobiotics biodegradation and metabolism \\ & Reaction module maps \\ & Chemical structure transformation maps \\
\hline
{Genetic Information Processing} & Transcription, Translation  \\ & Folding, sorting and degradation \\ & Replication and repair \\
\hline
{Environmental Information Processing} & Membrane transport, Signal transduction \\ & Signaling molecules and interaction \\ 
\hline
{Cellular Processes} & Transport and catabolism \\ & Cell motility, Cell growth and death  \\ & Cell communication \\
\hline
{Organismal Systems} & Immune system, Endocrine system  \\ & Circulatory system, Digestive system \\ & Excretory system, Nervous system \\ & Sensory system, Development \\ & Environmental adaptation \\
\hline
{Human Diseases} & Cancers, Immune diseases \\ & Neurodegenerative diseases  \\ & Substance dependence, Cardiovascular diseases \\ & Endocrine and metabolic diseases  \\ & Infectious diseases: Bacterial \\ & Infectious diseases: Viral \\ & Infectious diseases: Parasitic \\
\hline
\end{tabular}
\label{bigtab}
\end{table}

\newpage

\subsection*{Entropy of information flux distribution $\{S_F^i\}$}

We recall here the entropy-like measure $\{S_F^i\}$  proposed in \cite{Severini}. This measure is a single-node entropy measure, in fact, $\{S_F^i\}$ is a collection of entropy values, one for each node of the designed network ($i=1,...N$ where N is the total number of nodes). In \cite{Severini} the method was applied to four cancer datasets, three of them (``Wang", ``Loi", ``Frid") also considered for testing our entropy measure $S_{NE}$. We want to investigate how $\{S_F^i\}$ behaves both in cancer analysis and in ageing analysis,  using the same datasets considered for our entropy measure $S_{NE}$.\\
We compute $\{S_F^i\}$ using the following recipe:
\begin{enumerate}
\item choose the number of samples used to compute the correlation matrix $\{ c_{ij} \}$ ($R$ Pearson correlation coefficient) of our gene-expression profiles 
\item rescale the correlation values $c_{ij}$ between 0 and 1 ( $d_{ij}=(1+c_{ij})/2$))
\item compute the Hadamard product between the PPI and the distance matrix
\item for any given node $i$ with neighbors $j$ we can then assign a probability
distribution as follows
\begin{equation}
p_{ij}=\frac{d_{ij}}{\sum_j d_{ij}}
\end{equation}
\item The proposed entropy measure is the entropy of information flux distribution: it quantifies the amount of randomness/disorder of the local flux distribution surrounding any given node $i$
\begin{equation}
S_F^i=-\frac{1}{ \log k_i} \sum_j p_{ij} \log p_{ij}
\end{equation}
\item At the end we obtain, for each pool of samples, a vector of entropy values $\{S_F^i\}$.
\end{enumerate}
This measure is similar to the local disparity-heterogeneity index of weighted networks. When the flux is constrained along one direction $S_F^i=0$, while when the flux is equally distributed among all neighbors ($p_{ij}=1/k_i$) $S_F^i=1$. This perspective is quite different from $S_{NE}$ measure: for a specific network we have just a unique scalar entropy value instead of a vector of values, one for each node. The network is then considered as a whole system, not divisible in its single components.\\
In \cite{Severini} there was a comparison of local entropies for the proteins, between the non-metastatic (non-relapsing) and metastatic (relapsing) networks, in the three datasets. They restricted the analysis to proteins of degree $\ge10$ and because of the greater number of non-metastatic samples compared to the metastatic samples,  they considered metastatic and non-metastatic PPI-mRNA networks with the same number of samples, randomly choosing a non-metastatic subset. For each cancer dataset the performed 10 bootstraps of the non-metastatic samples to test the robustness of their results. The proposed measure always showed a small yet significant increase in entropy for the metastatic network (a Wilcoxon
paired test was performed).\\
We reproduce here the same analysis for ``Wang", ``Loi" and ``Frid", with 100 bootstraps. We also present the results for the subnetworks that survived to a Student's T test for uncoupled samples over the two different groups, with a $P=0.05$ significance threshold. The results are presented in Tab. \ref{tab_cancer_sev}. The majority of the results agrees with what was shown in \cite{Severini}, even if, with the selected subnetworks, we found some significant results with the opposite trend.\\
\begin{table}[h!]\small
\caption{Analysis of $\{S_F^i \}$ for the cancer datasets with 100 bootstraps: L0=Not Rel/Not Met and L1=Rel/Met (median of the entropy distribution). For each case we give the percentage of the bootstraps that showed the result, and also the percentage of the significant results among them (P$\le$ 0.05 with a Wilcoxon paired test). } 
\centering
\begin{tabular}{ |l|l|l|l|l|l|l|}
\hline
Dataset & proteins with $k\ge10$ & L1 & $\%$ L0 $<$ L1 & $\%$ signif. & $\%$ L0 $\ge$ L1 & $\%$ signif. \\
\hline
Wang & 1857 & 0.9973 & 100 & 100 & 0 & 0 \\ 
\hline
Wang 5$\%$ & 147 & 0.9967 & 100 & 100 & 0 & 0\\
\hline
Loi & 1857 & 0.9910 & 97 & 96.9 & 3 & 100\\ 
\hline
Loi 5$\%$ & 67 & 0.9928 & 83 & 21.7 & 17 & 23.5  \\
\hline
Frid & 1849 & 0.9932 & 98 & 100 & 2 & 100\\
\hline
Frid 5$\%$ & 19 & 0.9903 & 77 & 19.5 & 23 & 0\\
\hline
\end{tabular}
\label{tab_cancer_sev}
\end{table}
If we consider the whole Not Rel/Not Met datasets,  without performing bootstraps, we obtain a  different result,  since the Not Rel/Not Met entropy is always significantly higher than the Rel/Met entropy, except for Wang 5$\%$ (see Tab. \ref{tab_cancer_whole}).

\begin{figure}[h!]
\centering
\includegraphics[width=130 mm]{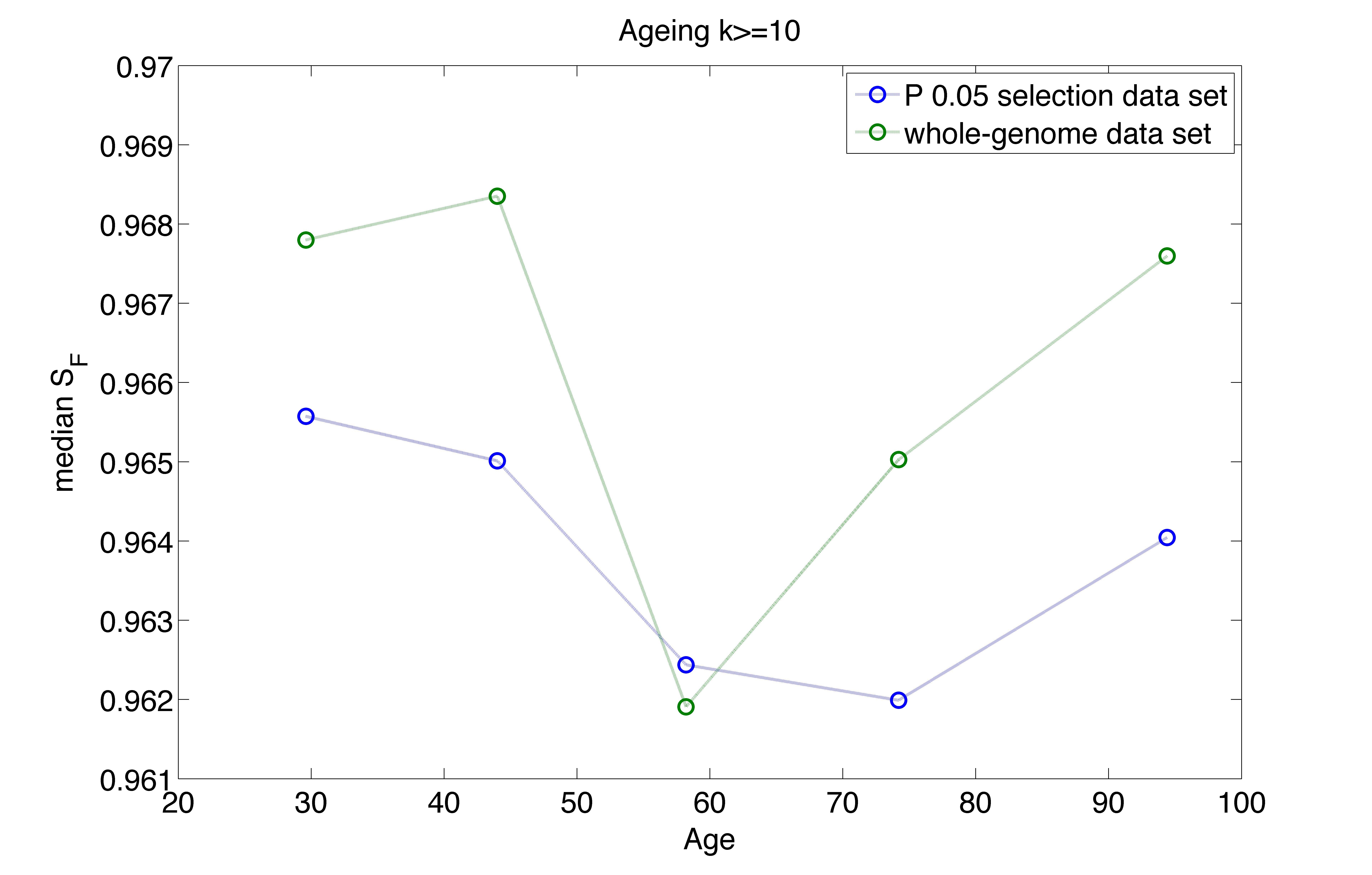}
\caption{Median values of $\{S_{F}^i\}$ for the whole-genome dataset (green) and for the $5\%$  selection subset (blue). }\label{severiniRageing}
\end{figure}

\begin{figure}[h!]
\centering
\includegraphics[width=130 mm]{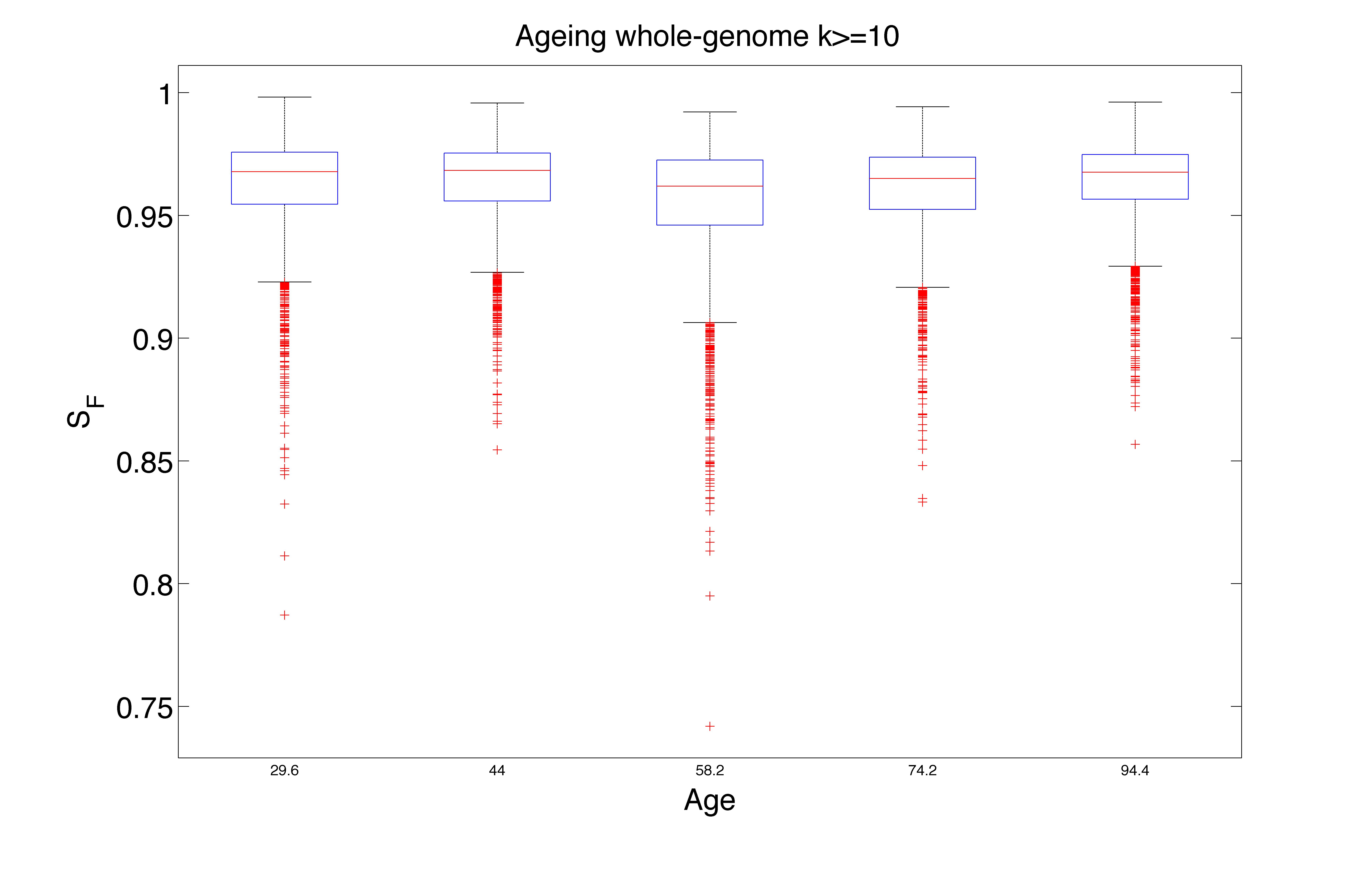}
\caption{Boxplot for $\{S_{F}^i\}$ of the whole-genome dataset.}\label{boxplotSeverini6353nodesv10}
\end{figure}

\begin{figure}[h!]
\centering
\includegraphics[width=130 mm]{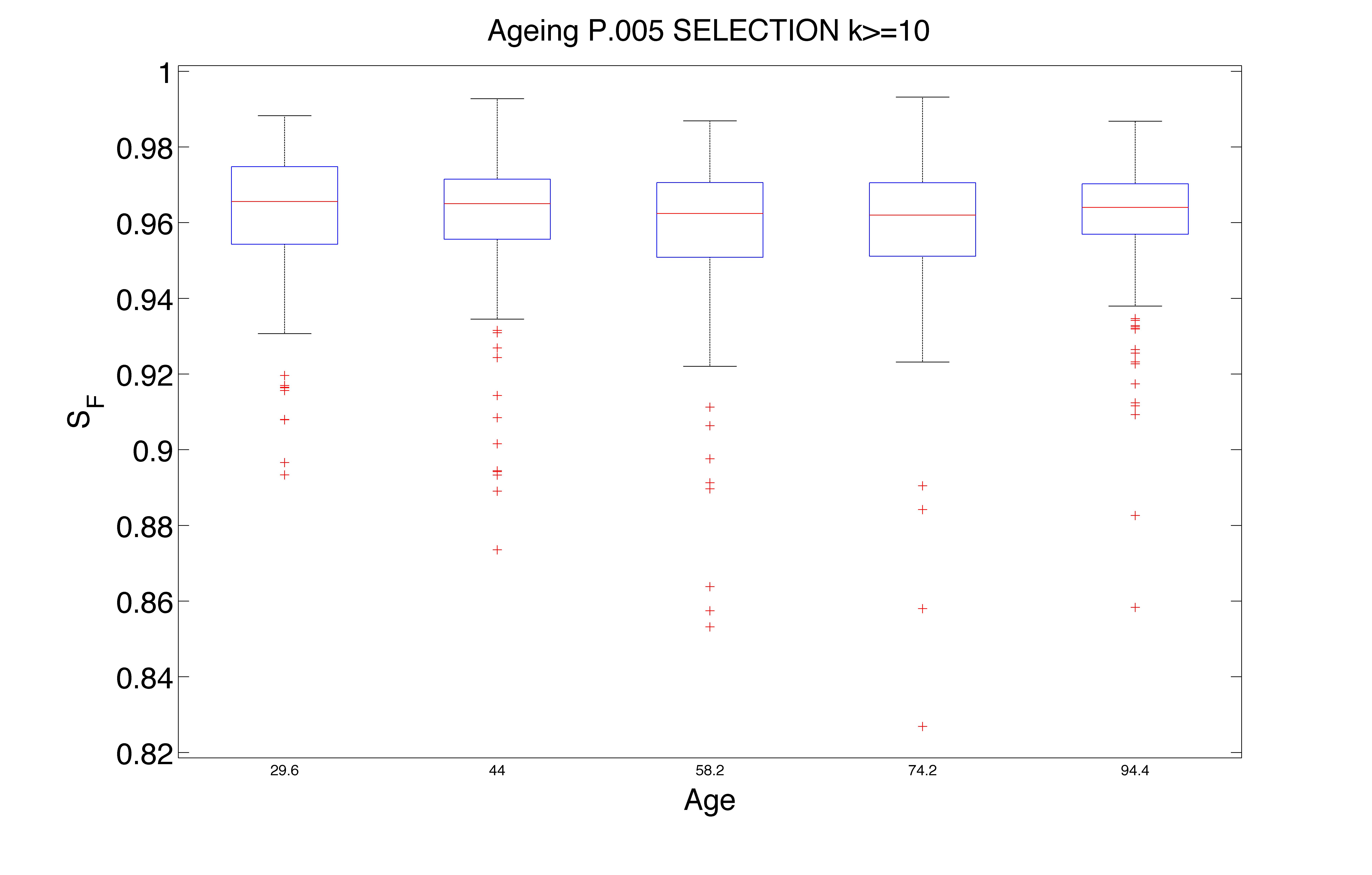}
\caption{Boxplot for $\{S_{F}^i\}$ of the $5\%$  selection dataset.}\label{boxplotSeveriniR638nodesv10}
\end{figure}

\begin{table}[h!]\small
\caption{Analysis of $\{S_F^i \}$ for the whole cancer datasets (without bootstrap): L0=Not Rel/Not Met and L1=Rel/Met. The P-value is given from the Wilcoxon paired test.}
\centering
\begin{tabular}{ |l|l|l|l|l|l|l|}
\hline
Dataset & proteins with $k\ge10$ & median L0 & median L1 & P-value \\
\hline
Wang & 1857 & 0.9975 & 0.9973 & 6.4 $\cdot 10^{-12}$ \\ 
\hline
Wang 5$\%$ & 147 & 0.9965 &  0.9967 & 0.03 \\
\hline
Loi & 1857 & 0.9928 & 0.9910 & 2.1 $\cdot 10^{-113}$ \\ 
\hline
Loi 5$\%$ & 67 & 0.9944 & 0.9928 & 1.8 $\cdot 10^{-7}$ \\
\hline
Frid & 1849 & 0.9952 & 0.9932 & 4.5 $\cdot 10^{-138}$\\
\hline
Frid 5$\%$ & 19 & 0.9944 & 0.9903 & 0.0055\\
\hline
\end{tabular}
\label{tab_cancer_whole}
\end{table}
We consider now the Ageing dataset and we study the distributions of $\{S_F^i \}$ for proteins with connectivity $\ge$10. We use both the whole-genome network of 6353 nodes (2972 nodes with degree $\ge$10 ) and the $5\%$  selection subset of 638 nodes (170 nodes with degree $\ge$10 ) . For each network we computed 5 correlation matrices, one for each age group. In Fig. \ref{severiniRageing} we plotted the related median values of $\{S_{F}^i\}$. for the two datsets. In Fig. \ref{boxplotSeverini6353nodesv10} and Fig. \ref{boxplotSeveriniR638nodesv10} we show their boxplots.
 
Applying the Kruskal-Wallis test we found $P=0.1315$ for the $5\%$  selection and $P=2\cdot 10^{-50}$ for the whole-genome network, thus we can see how significance is strongly dependent on node number and not on the number of samples, even if the gene (node) subset comes from a selection based on statistical significance test.

\end{document}